# GRASP MRI: A Decade of Innovation from Bench to Bedside

Li Feng, Kai Tobias Block, Hersh Chandarana, Daniel K Sodickson

1 Bernard and Irene Schwartz Center for Biomedical Imaging, Department of Radiology, New York University Grossman School of Medicine, New York, NY, USA.

2 Center for Advanced Imaging Innovation and Research ($CAI^2R$), Department of Radiology, New York University Grossman School of Medicine, New York, NY, USA.

Address correspondence to:

Li Feng, PhD

Center for Advanced Imaging Innovation and Research ($CAI^2R$)

New York University Grossman School of Medicine

660 First Avenue

New York, NY, USA 10029

Email: Li.Feng@nyulangone.org

## Abstract

GRASP (Golden-angle RAdial Sparse Parallel) MRI has emerged as one of the most influential motion-robust dynamic MRI frameworks over the past decade. By combining continuous golden-angle radial sampling with compressed sensing and parallel imaging, GRASP enables free-breathing data acquisition with flexible retrospective image reconstruction. Since its original introduction, the framework has evolved substantially and has inspired a broad range of technical developments, including motion-resolved reconstruction, real-time imaging, quantitative MRI, deep learning-enabled reconstruction, and multidimensional cardiovascular imaging. These advances have further expanded the role of GRASP MRI in a range of clinical applications where conventional breath-hold imaging is challenging. This review summarizes the technical evolution and clinical translation of GRASP MRI over the past decade, with a particular focus on the conceptual advantages of continuous radial acquisition, flexible retrospective reconstruction, and motion-robust imaging. Emerging developments in deep learning reconstruction, real-time volumetric imaging, and quantitative free-breathing MRI are also discussed together with future directions of motion-robust MRI acquisition and reconstruction.

## Introduction

Magnetic resonance imaging (MRI) is a powerful and versatile imaging modality that plays an essential role in clinical diagnosis. It offers excellent soft-tissue contrast, multiparametric information, and flexible imaging protocols, enabling comprehensive evaluation of a wide range of diseases without exposing patients to ionizing radiation. Nevertheless, the relatively slow imaging speed of MRI remains a significant challenge compared to other imaging modalities, making MRI exams susceptible to motion artifacts and limiting the applicability for dynamic imaging applications. Long acquisition times also necessitate breath-holding during imaging of moving organs such as the liver and the heart, which can be difficult or impractical for many patients. These challenges together have resulted in an MRI workflow that remains complex and cumbersome, even after decades of routine clinical use.

Over the past few decades, major advances in fast imaging techniques have been made to address these limitations, and these efforts have led to remarkable improvement in image quality, acquisition speed, and overall diagnostic performance. Among these developments, Golden-angle RAdial Sparse Parallel (GRASP) MRI stands out as a rapid, motion-robust dynamic imaging approach that enables free-breathing acquisition without the need for breath-holding (1). GRASP combines compressed sensing and parallel imaging with golden-angle radial sampling into a unified framework. This combination entails not only faster imaging but also establishes a new paradigm of rapid, continuous MRI with improved motion robustness, greater workflow efficiency, and enhanced flexibility for clinical use (2).

Since its introduction in 2012, GRASP MRI has been widely adopted both in research and clinical settings and has been applied across a variety of organ systems (3), including the brain (4), neck (5), breast (6), liver (7), kidneys (8), bowel (9), prostate (10), and bladder (11). Since 2017, GRASP MRI is also commercially available on Siemens MRI systems with FDA clearance for routine diagnostic use. This progress reflects not only the technical strengths of GRASP MRI but also its adaptability to real-world clinical workflows, particularly for patients with limited breath-hold capacity (12,13). Over the past decade, the original GRASP technique has also evolved into multiple advanced versions, extending the capabilities towards motion-resolved reconstruction (14), time-resolved 4D

MRI (15), real-time motion tracking (16), multiparametric mapping (17), and deep learning-based reconstruction (18). These new developments have broadened the scope and clinical impact of GRASP MRI from traditional diagnostic imaging to emerging applications in image-guided treatment.

The goal of this article is to present an overview of GRASP MRI, covering its technical foundation, major advances, and clinical applications. We also share our decade-long journey at NYU with GRASP MRI, including the history of its inception, development, and successful translation into routine clinical use. The review begins with a historical overview that outlines the motivation and early development of GRASP MRI, followed by a detailed summary of the technical framework, clinical implementation and impact. The subsequent two sections then highlight different methodological extensions and variants of GRASP MRI and their applications. Finally, we conclude the review with a discussion of current limitations of this technique and its future directions. By the end of the review, we hope to provide readers with a clear understanding of how GRASP MRI works, what has been achieved, and where this technique is headed next.

## GRASP MRI: A Historical Overview

The origins of GRASP MRI go back to the year 2010, when researchers at NYU were pursuing two separate but complementary paths of investigation that ultimately converged into what became the GRASP technique. The first direction focused on combining compressed sensing with parallel imaging to achieve highly accelerated dynamic MRI using undersampled Cartesian k-space trajectories. At that time, compressed sensing was still a relatively new concept but had quickly gained substantial attention in the MRI community (19–21). The NYU team was among the early groups to demonstrate that integrating compressed sensing with parallel imaging in a SENSE (sensitivity encoding) reconstruction framework could outperform either approach alone for accelerated dynamic imaging (22). This reconstruction strategy was successfully applied to several applications, including myocardial perfusion MRI (22,23), real-time cardiac cine MRI (24–26), phase-contrast cine MRI (27), and quantitative MR parameter mapping (28). These early efforts in combining compressed sensing and parallel imaging laid the groundwork for what later evolved into GRASP MRI and has since become standard in modern iterative MRI reconstruction.

The second direction centered on evaluating a fat-saturated, T1-weighted stack-of-stars 3D gradient echo (GRE) sequence called Radial VIBE (now known as StarVIBE), which was developed by Siemens in 2010 for motion-robust imaging (29). While the concept of stack-of-stars for radial sampling had been introduced in the late 1990s (30), it was not broadly available on commercial MRI scanners with vendor support until the release of Radial VIBE. As an extension of Siemens' VIBE sequence family, Radial VIBE employs a hybrid radial-Cartesian trajectory, implementing radial sampling for in-plane acquisition and Cartesian sampling along the slice encoding direction (31), as shown in Figure 1. The sequence was released as a work-in-progress (WIP) package on the Siemens MRI platform, with thorough optimization of key technical elements such as gradient delay correction and fat suppression, both of which are essential for routine clinical use. In the same year, NYU became the first academic institution to adopt Radial VIBE for clinical patient studies, and preliminary results were published the following year (32). The main motivation for using this sequence in the study was to leverage its motion robustness for free-breathing, contrast-enhanced multiphase liver MRI.

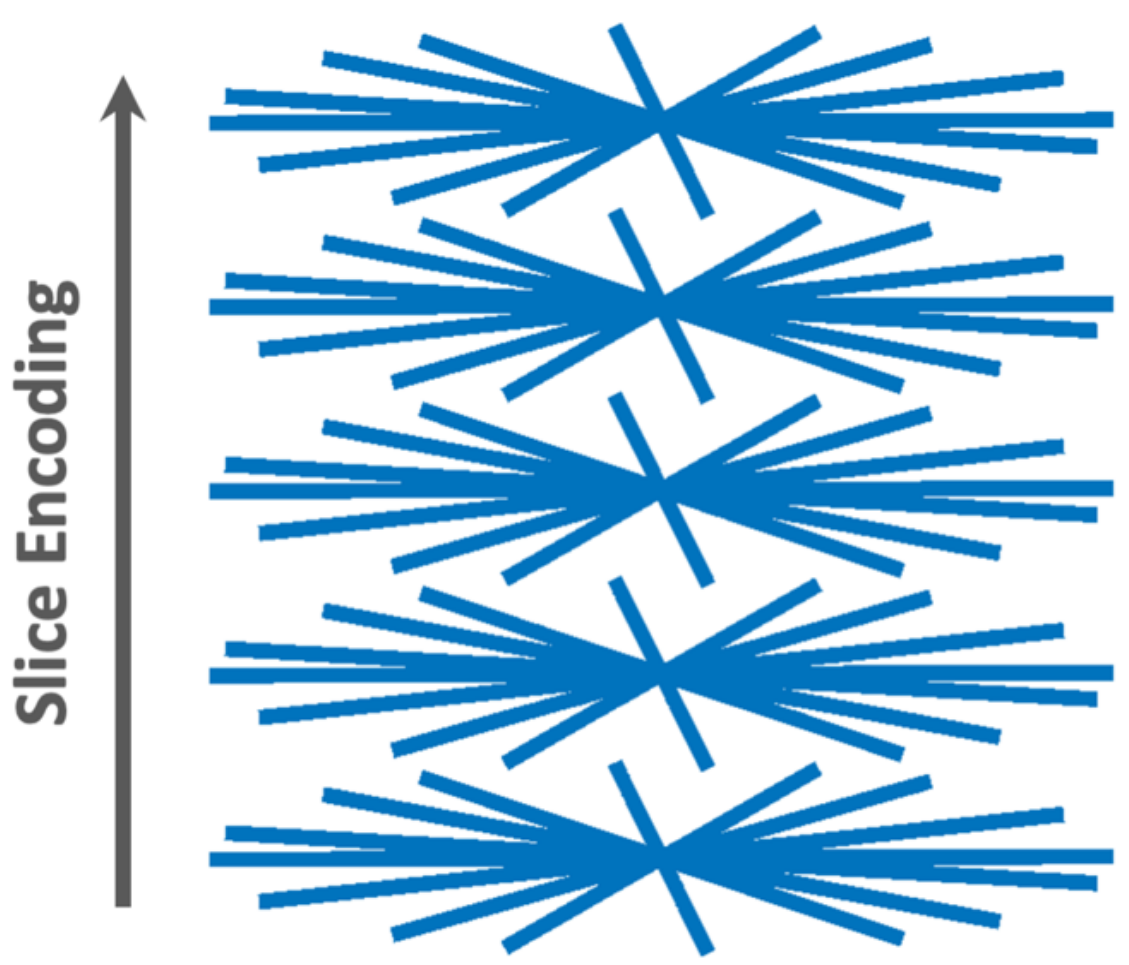

**Figure 1:** Stack-of-stars sampling trajectory. Radial sampling is used for in-plane encoding, while Cartesian encoding is applied along the slice direction. This hybrid strategy preserves the motion robustness of radial trajectories while maintaining key advantages of Cartesian-based slice encoding, such as robust fat suppression and compatibility with parallel imaging reconstruction.

By 2011, NYU investigators, including both radiologists and MRI physicists, began working together to integrate compressed sensing and parallel imaging with stack-of-stars sampling. Around this time, Kai Tobias Block, the lead developer of the Radial VIBE sequence, joined NYU as a faculty member, bringing strong expertise in radial imaging. The arrival of Tobias catalyzed the convergence of the previously separate research directions into a unified framework. Initially, the integration did not include golden-angle sampling, even though this was a built-in feature of the Radial VIBE sequence. However, the team quickly recognized that golden-angle sampling offers great potential for continuous data acquisition in accelerated dynamic imaging, enabling flexible retrospective reconstruction particularly well suited for dynamic contrast-enhanced MRI (DCE-MRI) (33,34). This eliminated the need to predefine the desired temporal resolution or specify in advance how each contrast-enhanced phase should be acquired. Instead, all reconstruction parameters could be defined retrospectively after data acquisition. For example, in DCE-MRI, the same dataset can be reconstructed with a lower temporal resolution for qualitative clinical assessment or with a higher temporal resolution for quantitative perfusion analysis (6,35), as illustrated in Figure 2. The name GRASP MRI was selected in 2012 during a brainstorming session, and the team was enthusiastic about its potential to transform clinical workflows by enabling free-breathing, high-resolution dynamic imaging that combines speed, flexibility, and motion robustness within a single framework.

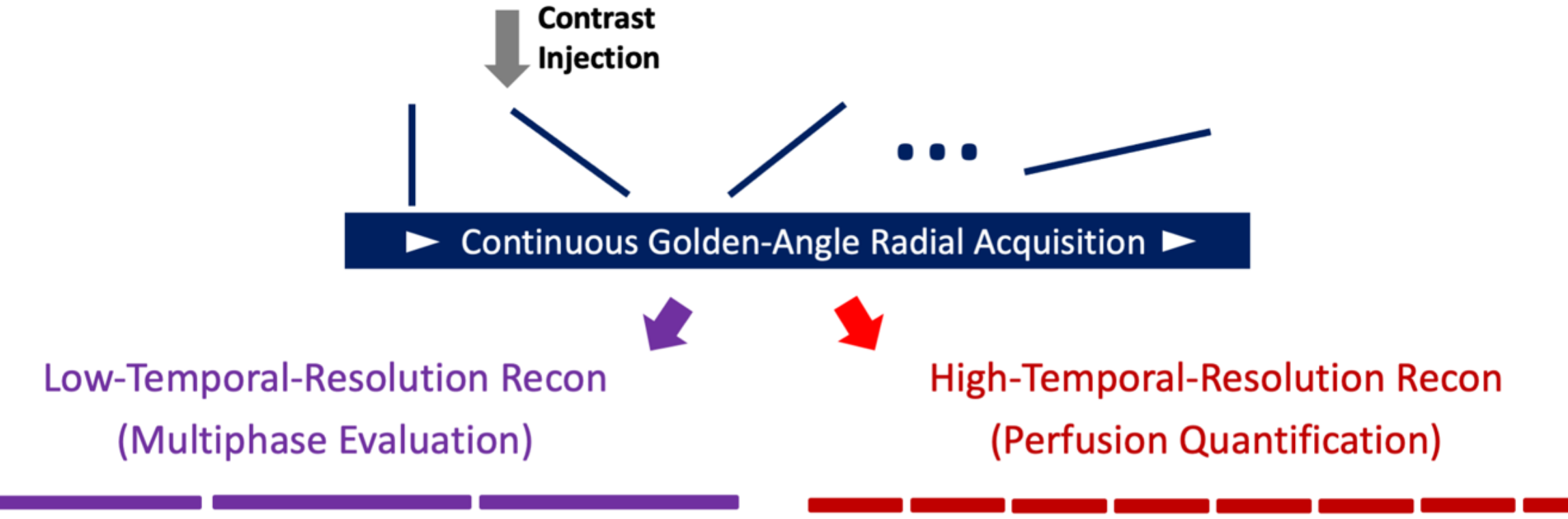

**Figure 2:** Flexibility of golden-angle radial sampling. The continuously rotating golden-angle acquisition enables retrospective reconstruction at different temporal resolutions for tailored clinical needs, such as low–temporal-resolution images for conventional multiphase assessment and high–temporal-resolution series for perfusion quantification.

## Technical Components of GRASP: A Link to the Past

GRASP MRI integrates multiple technical components spanning both data acquisition and image reconstruction. On the acquisition side, it employs a golden-angle radial sampling scheme, initially implemented using the stack-of-stars trajectory and later extended to other radial sampling schemes. On the reconstruction side, GRASP combines compressed sensing and parallel imaging with spatiotemporal regularization to recover dynamic images from undersampled data. In fact, these acquisition and reconstruction strategies were previously proposed independently in earlier works, but their synergy in GRASP ultimately enabled a practical, rapid, and motion-robust solution for dynamic MRI. This section revisits these components from a historical perspective and highlights how their integration provides key advantages in imaging speed, motion robustness, and reconstruction flexibility.

### Golden-Angle Radial Sampling

Radial sampling has a long and influential history and was, in fact, the first data acquisition strategy demonstrated by Paul Lauterbur in his pioneering MRI experiments (36). With the advent of Fourier imaging, however, Cartesian sampling quickly became the standard because of its simplicity and better tolerance to system imperfections. Clinical adoption of radial sampling was initially very limited for several reasons (31). First, since data are collected on a non-Cartesian grid, image reconstruction from radial data requires computationally intensive algorithms, such as the “gridding” method. Second, because each view in radial sampling is acquired at a different angle (referred to as “spoke), the method is sensitive to gradient delays and off-resonance effects. Third, repeated sampling of the k-space center in radial MRI reduces the effectiveness of fat saturation and leads to undesired contrast mixing, which hinders its use in certain applications such as fast spin-echo (FSE) imaging.

While radial sampling remained primarily a research tool for decades, important advances have expanded its capabilities and clinical translation. As early as the 1990s, studies already demonstrated that radial trajectories offer greater motion robustness than Cartesian sampling (37,38). In the meantime, sliding-window or view-sharing reconstruction strategies were developed for continuous data acquisition in real-time dynamic imaging (39–41). Radial sampling has since been applied not only to standard T1-weighted acquisition but also to T2-weighted MRI (42), GRASE (gradient and spin echo) imaging (43), quantitative parameter mapping (44,45), ultrashort echo time (UTE) imaging (46,47), and other specialized applications (48–50). The concept of stack-of-stars sampling, which is widely adopted today, was initially proposed for contrast-enhanced MR angiography by Peters et al. in the late 1990s to balance motion robustness and acquisition efficiency (30).

The concept of golden-angle radial sampling was first introduced at the 2005 ISMRM Annual Meeting, where it was demonstrated by Winkelmann et al. for single-shot T1 mapping with a Look-Locker sequence (51). This work was soon expanded into a full publication in 2007, which described how golden-angle ordering optimizes k-space coverage for time-resolved MRI (33). The key advantage of golden-angle sampling lies in its flexibility, which allows a single dataset to be retrospectively reconstructed at multiple temporal resolutions to address different clinical questions. At the 2006 ISMRM Annual Meeting, well before the development of GRASP, the feasibility of applying this idea to dynamic imaging had been demonstrated for DCE-MRI using view-sharing reconstruction (34). In addition, the flexibility of golden-angle radial sampling allows retrospective sorting of acquired spokes according to underlying respiratory or cardiac states (52), an idea later implemented in the eXtra-Dimensional GRASP (XD-GRASP) technique (14).

Golden-angle radial sampling has since been refined in two main directions that continue to be widely explored. The first was the introduction of the tiny golden-angle scheme (53), designed to minimize eddy-current artifacts sometimes observed in 2D radial imaging due to the large gradient jumps between adjacent spokes (54). The second extended golden-angle sampling to true 3D radial acquisitions based on the Koosh-ball pattern, from which two notable rotation strategies were proposed: one derived from 2D golden means (55) and the other from spiral phyllotaxis (56). These 3D variants enable isotropic coverage and spatial resolution, which have proven particularly valuable in cardiovascular and lung MRI (57–60).

Among the various radial sampling schemes, stack-of-stars has been the most widely adopted on modern MRI scanners. There are several reasons for the widespread adoption (31). First, the hybrid sampling scheme provides an effective balance between motion robustness and spatial encoding efficiency. Radial in-plane sampling offers improved tolerance to motion artifacts compared to Cartesian sampling, while Cartesian encoding along the slice direction simplifies reconstruction compared to full 3D radial imaging. This is particularly advantageous in applications like abdominal MRI, where isotropic spatial resolution and volumetric coverage are not always necessary. Second, Cartesian encoding along the slice direction supports effective fat suppression with conventional methods, which is essential for many routine clinical protocols. This implementation also enables reconstruction of different image slices in parallel after disentangling the slice dimension using a one-dimensional FFT, thereby facilitating faster reconstruction. These practical benefits, along with technical improvements in hardware design, gradient-delay correction, and robust vendor support, have made radial sampling clinically viable. The Radial VIBE sequence from Siemens represented the first implementation of stack-of-stars acquisition by a major vendor, and it has since been widely deployed in routine practice.

**Radial MRI Reconstruction and the Synergy with Compressed Sensing**

Before the advent of compressed sensing, reconstruction from radial MRI data primarily relied on relatively straightforward methods such as gridding and view sharing. View sharing, in particular, was widely explored for dynamic radial MRI reconstruction, as it allowed acquired spokes to be retrospectively grouped into overlapping temporal frames to achieve high frame rates, albeit with a potential risk of temporal blurring. While standard view-sharing reconstruction was conceptually simple and computationally efficient, it did not fully address the challenge of undersampling artifacts. Improved techniques, such as k-space weighted image contrast (KWIC) (61) and highly constrained projection reconstruction (HYPR) (50), were later developed to address these challenges, but various limitations complicated their practical use, and these methods did not achieve widespread clinical adoption.

In the early 2000s, several years after parallel imaging was proposed for accelerated MRI (62–64), different research groups began to adapt the concept for radial acquisition (65–67). However, the use of parallel imaging, whether through image-domain methods such as SENSE or k-space-domain methods such as GRAPPA (Generalized Autocalibrating Partially Parallel Acquisitions), has been primarily confined to Cartesian MRI, and clinical adoption of non-Cartesian parallel imaging has remained limited. This is largely due to the more complex reconstruction process and substantially higher computational demand for inverting large, ill-conditioned encoding matrices (68). A practical exception has been the application along the slice-encoding direction in stack-of-stars imaging. As early as 2005, GRAPPA was successfully combined with stack-of-stars sampling to accelerate slice encoding (69). Reconstruction was performed by first applying GRAPPA along the slice dimension, followed by gridding for in-plane reconstruction.

A major shift occurred in the mid-2000s with the introduction of compressed sensing to MRI (19). At the 2005 ISMRM Annual Meeting, Lustig and colleagues presented L1-constrained reconstruction for accelerated imaging, initially using randomly perturbed spiral undersampling to maximize incoherence (70). At the same meeting, Velikina demonstrated the use of spatial total variation (TV) constraints for reconstructing undersampled MRI data (71). In 2006, Lustig et al. expanded these ideas to variable-density Cartesian undersampling (72) and introduced the k-t SPARSE framework for dynamic MRI (73). In the same year, researchers from Siemens Corporate Research applied L1 constraints to undersampled radial MRI in phantom studies, marking the first demonstration of compressed sensing for radial trajectories (74).

In 2007, Lustig et al. published the landmark “Sparse MRI” paper (19), now a foundational work in the field. In the same year, two groups independently demonstrated the combination of compressed sensing with radial sampling. In the first work, Ye et al. used the Focal Underdetermined System Solver (FOCUSS) algorithm to iteratively solve for sparse solutions (75), while Block et al. applied a spatial TV constraint for image reconstruction from undersampled radial MRI data (21). Researchers also recognized the potential combination of compressed sensing with parallel imaging to further improve reconstruction performance (22,76,77). This synergy arises for two main reasons. First, the coil-sensitivity encoding in parallel imaging helps to suppress aliasing artifacts from undersampling, thereby facilitating more effective compressed sensing reconstruction. Second, the L1 constraint in compressed sensing helps to mitigate noise amplification in parallel imaging, particularly at high acceleration rates. The work presented by Block et al. in 2007 also proposed to incorporate parallel imaging into radial compressed sensing reconstruction using the SENSE framework, which is generally considered the first demonstration of this combination (21). In 2008, Liu et al. proposed SparseSENSE, which combined compressed sensing and SENSE for Cartesian sampling (77). Soon after that, Otazo et al. extended this method to k-t SPARSE-SENSE for dynamic Cartesian MRI with a time-varying, variable-density random undersampling scheme (22).

In addition to specific sampling strategies, compressed sensing reconstruction requires a sparsifying transform. Early compressed sensing reconstruction for dynamic MRI focused on two options. The first approach, initially implemented in the k-t SPARSE framework, used a temporal fast Fourier transform (FFT) to exploit temporal sparsity (73). The use of temporal FFT for dynamic MRI reconstruction had been introduced earlier by Tsao et al. in their k-t BLAST (Broad-use Linear Acquisition Speed-up Technique) method in 2003 (78), though not in the context of compressed sensing. The second approach, demonstrated by Adluru et al. for both Cartesian and radial MRI, enforced a temporal TV constraint for dynamic compressed sensing MRI reconstruction (79,80). While both approaches proved to be effective, a study in 2013 by Feng et al. compared the two temporal sparsity constraints and found that temporal TV outperformed temporal FFT for dynamic compressed sensing MRI reconstruction (24). Since then, temporal TV has become one of the commonly used sparsifying transforms in iterative dynamic MRI reconstruction, including GRASP MRI.

At the 2012 ISMRM Annual Meeting, the NYU team presented GRASP MRI that combined compressed sensing and parallel imaging with stack-of-stars golden-angle radial sampling, and demonstrated its performance in various motion-robust DCE-MRI applications (81). In many respects, GRASP MRI represented a natural evolution of earlier efforts, particularly as a synergy between the static radial MRI reconstruction framework of Block et al. and the k-t SPARSE-SENSE approach (1). More importantly, GRASP MRI demonstrated for the first time that rapid, continuous golden-angle radial acquisition could enable free-breathing body MRI with retrospectively selectable temporal resolution. This innovation overcame major limitations of earlier techniques and established a new paradigm for dynamic imaging with great flexibility.

## Clinical Translation and Impact

The successful translation of GRASP into clinical practice has been a major factor contributing to its broad impact. This achievement has distinguished GRASP MRI from many other dynamic MRI reconstruction techniques developed over the past decades, as bringing new reconstruction methods into routine use often falters due to obstacles at multiple levels. This section summarizes the collaborative efforts between NYU and Siemens teams that made this transition possible.

### Workflow Integration in Collaboration with Siemens

The first GRASP reconstruction pipeline was implemented in MATLAB, using spatiotemporal TV constraints with separate parameters controlling the balance between spatial and temporal regularizations. At that time, however, several barriers hindered the clinical translation of GRASP MRI.

First, iterative reconstruction is typically computationally demanding. For example, the initial MATLAB implementation of GRASP required ~15-30 minutes to reconstruct a single slice. Second, even if image reconstruction could be accomplished in a reasonable timeframe, integrating it into the MRI scanner workflow and automatically transferring the resulting images to the PACS system posed a significant challenge. Overcoming these challenges required a sustained team effort, driven not only by technical innovation but also by strong support from the NYU leadership, collaborating radiologists, and the IT department, whose contributions were essential at every stage. Together, these efforts established the groundwork for subsequent improvements and optimizations in partnership with Siemens.

To address the computational bottleneck, Robert Grimm, then a PhD student jointly affiliated with the Friedrich-Alexander University of Erlangen-Nuremberg and Siemens, visited NYU for six months in 2012. He re-implemented the entire GRASP reconstruction pipeline in C++, which substantially accelerated image reconstruction. With parallel computing, reconstruction time for an entire 3D volume was reduced to under 30 minutes, compared to 15-30 minutes per slice in MATLAB. This marked the first major step toward practical clinical translation.

The challenge of clinical translation was addressed in 2013 by Kai Tobias Block, who developed the Yarra framework (82), a software tool that has been used at NYU since then. As shown in Figure 3, Yarra connects the MRI scanner to an external reconstruction server and the PACS system. After a scan is completed, the technologist simply enters the accession number into Yarra on the scanner console. The MRI raw data are then automatically transferred to the external server, where the C++ GRASP pipeline reconstructs 4D dynamic images. Once reconstruction is done, Yarra automatically sends the images into PACS. This workflow is fully automated and requires only a single click on the scanner. In addition, Yarra also supports scheduled raw data transfer to outside servers (e.g., overnight when the scanner is idle). Over the past decade, this framework has proven to be a powerful solution for translating iterative reconstruction into routine clinical practice.

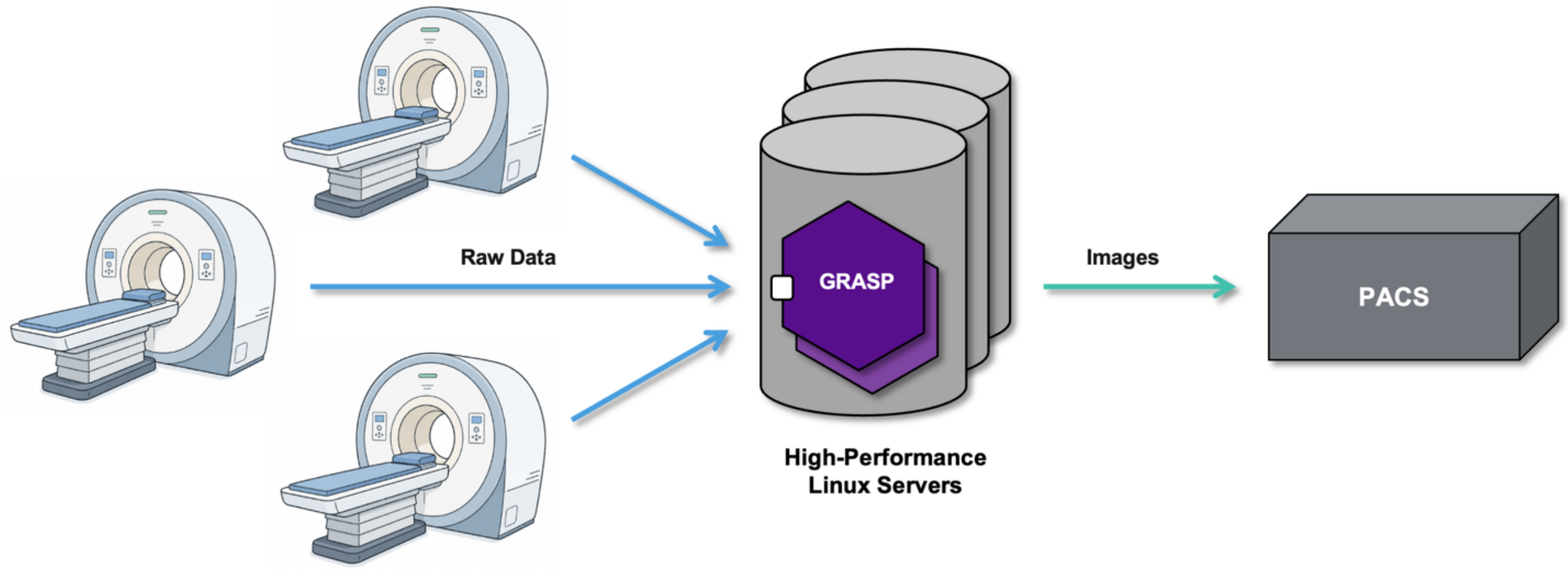

**Figure 3:** Workflow integration of GRASP MRI using the Yarra framework. After data acquisition, the MRI raw data are automatically transferred from the scanner to an external reconstruction server, where GRASP reconstruction generates 4D dynamic images. Once reconstruction is done, images are automatically sent into the PACS system.

Soon after GRASP was implemented and evaluated clinically at NYU, Siemens began integrating the technique directly into their MRI systems. After multiple years of optimization, GRASP MRI was formally introduced by Siemens as a product with FDA clearance at the 2017 European Congress of Radiology (ECR) under the name GRASP-VIBE. Since then, GRASP MRI has been available worldwide on the Siemens MRI platform and has been adopted by numerous clinical sites.

**Limitations of Standard GRASP MRI**

Despite its simplicity and robustness, the standard GRASP framework has several limitations that continue to affect its performance in clinical practice.

One issue is latency. Although Yarra has enabled the routine use of GRASP MRI, a 10-60-minute delay remains between the completion of a scan and the availability of reconstructed images for interpretation. While this latency does not usually affect non-emergent patient care, it reduces workflow efficiency. To assist technologists, time-averaged images are generated on the scanner immediately after acquisition to verify scan quality and contrast injection. However, while these images are useful for quick checks, they cannot replace the fully reconstructed dynamic image series required for diagnosis. Meanwhile, although this was further optimized in the Siemens GRASP-VIBE implementation with improved algorithms and integration, the slow reconstruction speed inherent to iterative methods remains a major bottleneck.

Another limitation involves imperfect bolus timing. In conventional contrast-enhanced multiphase MRI, a bolus timing step is typically required to ensure optimal capture of the arterial phase. GRASP MRI omits this step for simplicity, which can lead to suboptimal arterial phase timing. The flexibility of GRASP MRI to reconstruct images with shifted data sorting can mitigate this limitation, but this requires additional processing and cannot fully substitute for optimized prospective bolus timing.

A third challenge is the presence of residual streaking artifacts. In radial MRI, gradient imperfections and off-resonance effects often generate bright peripheral hotspots, which in turn give rise to strong streaking artifacts that propagate throughout the image (83). These artifacts are difficult to suppress completely with standard iterative reconstruction, and they may impact relevant anatomy and potentially reduce diagnostic confidence in certain scenarios.

Motion blurring is another limitation, especially in moving organs such as the liver. Although radial sampling is inherently more robust to motion than Cartesian sampling, it is not immune to motion effects (84). In radial MRI, motion causes image blurring rather than ghosting, which impairs the visualization of fine anatomical structures and dynamic contrast patterns. Such blurring can reduce the accuracy of detecting small lesions or subtle enhancement differences, which has, to some extent, limited the use of standard GRASP MRI in the upper abdomen for imaging the liver, pancreas, and surrounding structures.

Many of these challenges are not unique to GRASP MRI but are also encountered in other radial MRI techniques. Over the past decade, various solutions have been proposed to address these limitations, including several contributions from the NYU team, which will be discussed in detail in the next section. Briefly, long reconstruction times can be accelerated with deep learning-based methods (see DeepGRASP). Suboptimal bolus timing can be alleviated with contrast-guided data sorting (see RACER-GRASP) or by leveraging high-temporal-resolution reconstruction (see GRASP-Pro). Residual streaking artifacts can be reduced using an approach called “unstreaking”, while motion-related blurring can be mitigated through motion-resolved reconstruction (see XD-GRASP), adaptive data weighting (see RACER-GRASP), or sub-second temporal resolution reconstruction (see GRASP-Pro). Collectively, these solutions are expected to further improve the performance and clinical reliability of GRASP MRI.

## Variants of GRASP: Advances and Extensions

Since its original development, GRASP MRI has inspired a wide range of methodological extensions aimed at addressing various limitations and expanding its clinical and research applications. These variants build upon the GRASP framework through innovations in data acquisition, reconstruction strategies, and computational modeling, leading to improved spatiotemporal resolution, motion robustness, and flexibility across different organ systems and imaging needs. This section provides a brief overview of major GRASP variants developed by the NYU team over the past decade and highlights their potential applications.

### XD-GRASP

Although radial sampling offers improved robustness to motion, motion-induced blurring can still degrade image quality, particularly in patients with irregular or deep breathing. The XD-GRASP (eXtra-Dimensional GRASP) technique was developed to address this limitation by sorting continuously acquired radial spokes according to underlying motion information, which generates an additional motion-resolved dimension (14). The use of golden-angle radial sampling provides the flexibility to achieve adequate k-space coverage after motion-based data sorting, while compressed sensing reconstruction with a temporal sparsity constraint is applied to suppress undersampling artifacts caused by data binning. Compared with conventional motion correction approaches such as image registration, XD-GRASP provides more robust and effective motion management (85) and also yields additional motion information that may be of clinical value (86–89). This approach has demonstrated superior performance over standard GRASP in free-breathing liver DCE-MRI (90), cardiac imaging (14), and other motion-sensitive applications (85,86,89).

**GROG-GRASP**

In standard GRASP reconstruction, gridding is performed within each iterative step, which prolongs reconstruction time compared with iterative Cartesian reconstruction that relies only on standard FFT. GROG-GRASP incorporates the GRAPPA Operator Gridding (GROG) technique, originally proposed by Seiberlich et al. as an alternative to conventional gridding (91,92), to pre-shift radial data onto a Cartesian grid using coil sensitivity information (93). This preprocessing step applies parallel imaging to estimate k-space data on a Cartesian grid from nearby radial samples, enabling the subsequent iterative reconstruction to be performed entirely in Cartesian space. As a result, computationally expensive gridding operations are not needed, and the reconstruction efficiency can be improved.

**RACER-GRASP**

RACER-GRASP (Respiratory-weighted, Aortic Contrast Enhancement-guided and coil-unstReaking) addresses several limitations of GRASP collectively (94). First, it introduces a contrast-guided data sorting strategy to optimize arterial phase reconstruction. Second, it employs motion-weighted reconstruction (also known as soft-gating), which assigns different weights to radial k-space data based on their underlying respiratory states, thereby reducing contributions from unfavorable motion phases. Unlike XD-GRASP, which explicitly bins data into multiple motion phases, RACER-GRASP does not require full motion-resolved reconstruction and is advantageous in terms of efficiency. Third, RACER-GRASP incorporates an “unstreaking” algorithm to suppress residual streaking artifacts (95). By applying coil-wise soft weighting before iterative reconstruction, this method attenuates the contribution of coil elements prone to strong streaks. Together, these strategies improve arterial phase delineation, reduce motion blurring, and enable more effective suppression of streaking artifacts.

**3D Koosh-Ball GRASP**

While GRASP is commonly implemented with stack-of-stars sampling, it is not limited to this approach. True 3D radial sampling, such as the Koosh-ball trajectory, provides isotropic volumetric coverage and spatial resolution, which are advantageous for applications like cardiovascular imaging. Building on this concept, the NYU team and collaborators from the University of Lausanne jointly developed a free-running 5D whole-heart MRI framework, which combines golden-angle 3D radial Koosh-ball sampling for continuous data acquisition with XD-GRASP for cardiac- and respiratory-resolved reconstruction (96). This framework has since been extended to other applications, such as flow MRI (5D flow) (97) and 4D lung MRI (89), enabling motion-resolved volumetric imaging with promising clinical value.

**GRASP-Pro**

Standard GRASP reconstruction relies on spatiotemporal TV constraints for dynamic image reconstruction. GRASP-Pro improves this by incorporating a low-rank subspace model that more effectively captures temporal correlations across frames (98). In this framework, dynamic images are represented within a low-rank subspace by projection onto a pre-estimated temporal basis. This reduces the number of unknowns for the reconstruction, thereby improving both image quality and computational efficiency. This strategy also supports dynamic reconstruction at ultrahigh temporal resolution (15). For example, GRASP-Pro has been applied to 4D liver DCE-MRI with sub-second temporal resolution, which inherently resolves respiratory motion and therefore eliminates the need for additional motion compensation (99), as shown in Figure 4. It also enables simultaneous reconstruction of multiple arterial phases to fully remove the need for conventional bolus timing or contrast-guided data sorting. In addition to liver imaging, GRASP-Pro has been applied to DCE-MRI of the breast with sub-second temporal resolution to enable more accurate perfusion quantification (100).

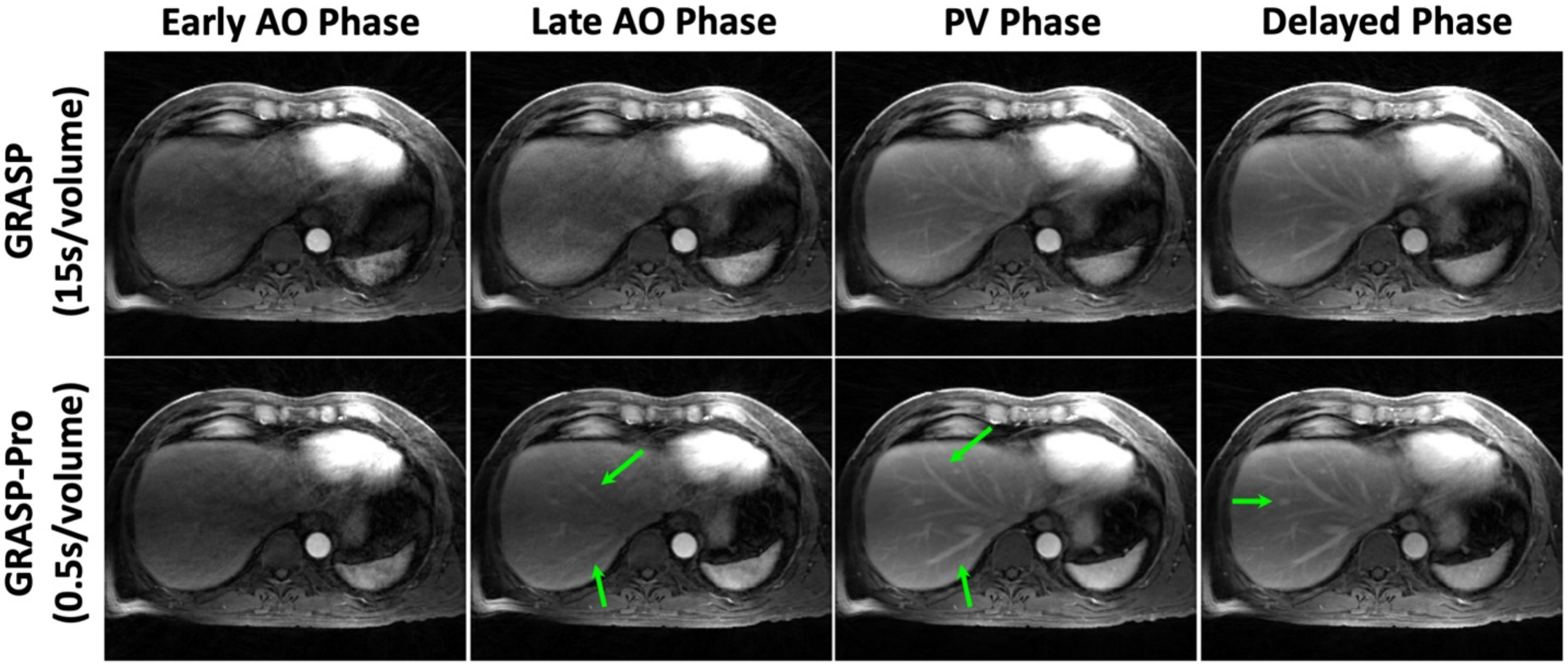

**Figure 4:** Comparison between standard GRASP MRI and GRASP-Pro with sub-second temporal resolution. Acquiring a full 3D volume within one second intrinsically resolves respiratory motion without the need for additional motion correction. Figure reproduced from *NMR in Biomedicine (2024 Dec;37(12):e5262)* with permission from the journal.

**MP-GRASP**

Standard GRASP MRI typically uses a steady-state acquisition without magnetization preparation. MP-GRASP (Magnetization-Prepared GRASP) incorporates magnetization-preparation pulses to provide additional contrast and enable improved contrast or quantitative parameter mapping (17). This approach facilitates several advanced applications, including free-breathing T1 mapping with inversion-recovery preparation (17,101,102), non-contrast dynamic 4D MRA using arterial spin labeling (ASL) (103), and free-breathing chemical exchange saturation transfer (CEST) MRI (104).

**Multi-Echo GRASP MRI**

In parallel with the development of different GRASP variants, stack-of-stars acquisitions were also adapted for multi-echo imaging by acquiring multiple echoes within each TR by the different research teams (105,106). This enables free-breathing fat/water separation and R2* estimation, which is particularly valuable for liver MRI exams, where separating fat and water signal is essential for assessing steatosis and other metabolic conditions. Beyond liver imaging, multi-echo stack-of-stars has also shown potential clinical value in breast MRI, where improved fat suppression enhances diagnostic performance (107). This imaging scheme was later combined with GRASP for DCE-MRI to enable simultaneous dynamic imaging and fat quantification at each contrast phase (108). More recently, multi-echo acquisition has also been incorporated into MP-GRASP for water-specific parameter quantification (17,109), a direction that is gaining increasing attention in liver imaging, where fat is known to be a confounding factor for quantitative measurements.

**Live-View GRASP**

Live-View GRASP is an extension of GRASP MRI for real-time image-guided interventions, such as MRI-guided radiation therapy (16). A major challenge in this application is the inherent latency of 3D MRI, arising from time-consuming data acquisition and image reconstruction, which often exceeds the clinical requirements for real time guidance. To address this challenge, several novel techniques, including Live-View GRASP, have been proposed (16,110–112). In Live-View GRASP, the imaging workflow is divided into two stages: an off-line or off-view stage and a live-view stage, as shown in Figure 5. During the off-line stage, free-breathing time-resolved 4D (3D + motion) images are acquired and reconstructed to form a motion-resolved image database, where each 3D image in the database is linked to a low-resolution 2D navigator that represents a specific respiratory state. During the live-view stage, only 2D navigators are acquired in real time, which can be rapidly matched to the off-line database for retrieving the best-matching 3D image. This imaging strategy shifts the computational burden to the off-line stage, which can be performed prior to treatment, while enabling fast and efficient live-view imaging during therapy.

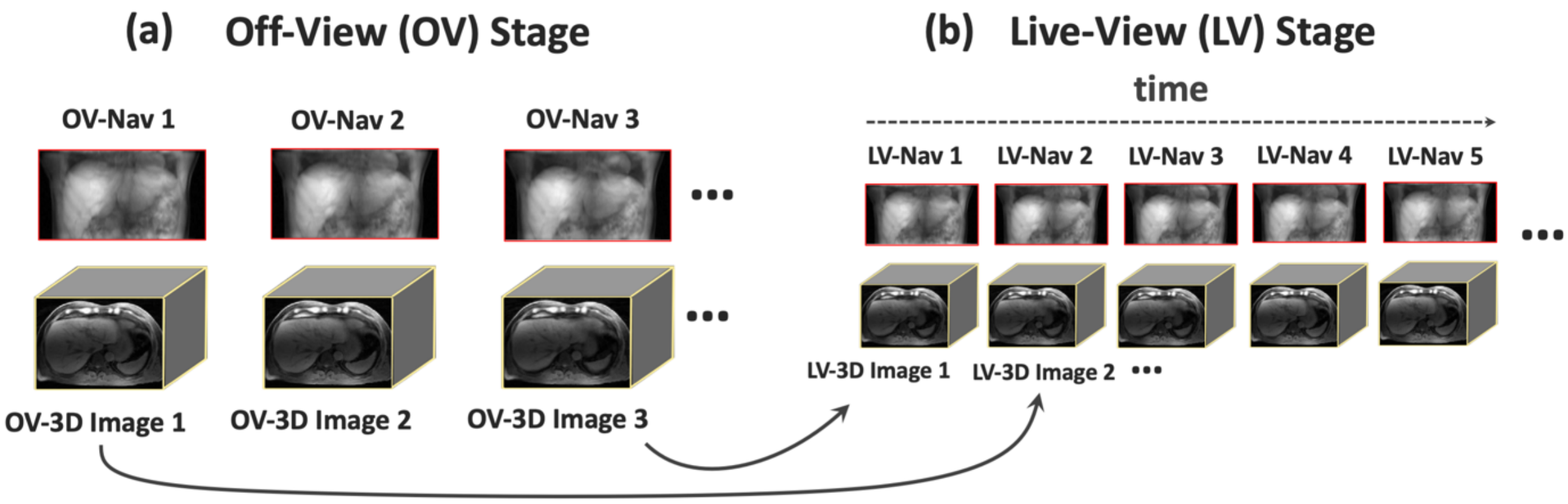

**Figure 5:** Live-View GRASP MRI framework. The workflow consists of two stages: an off-line (or off-view) learning stage and a live-view stage. In the off-line stage, free-breathing, time-resolved 4D (3D + motion) images are acquired and reconstructed to build a motion-resolved database, where each 3D image is linked to a corresponding low-resolution 2D navigator. During the live-view stage, only 2D navigators are acquired in real time and matched to the database to retrieve the best-corresponding 3D image with minimal latency. Figure reproduced from *Magn Reson Med. 2023 Sep;90(3):1053-1068* with permission from the journal.

**DeepGrasp**

DeepGrasp is a recent innovation that integrates self-supervised deep learning into the GRASP framework (18,113). Building on the low-rank subspace model of GRASP-Pro, DeepGrasp incorporates subspace modeling into a neural network-based reconstruction pipeline. This accelerates image reconstruction substantially while maintaining image quality, even when reconstructing large numbers of dynamic frames. Importantly, self-supervised learning in DeepGrasp avoids the need for fully sampled training data, which are not available from in-vivo GRASP scans. More recently, an all-in-one DeepGrasp model has been developed to generalize this technique across different organs, spatial resolutions, and temporal frames within a single trained model, with preliminary results shown in Figure 6. Such a unified approach holds the potential to streamline clinical deployment by reducing the need for retraining, while preserving the flexibility and robustness across different dynamic imaging applications.

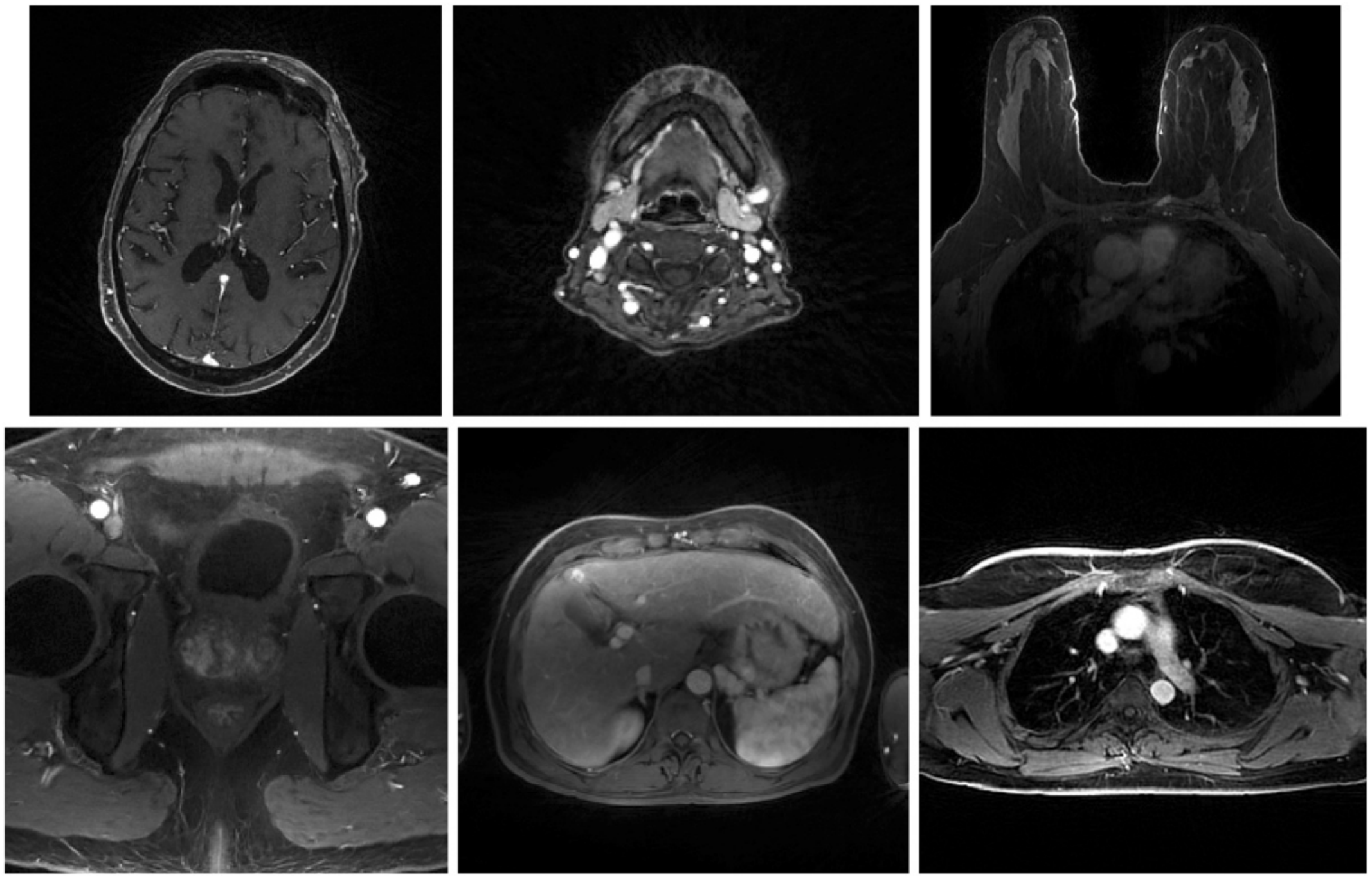

**Figure 6:** Preliminary results from a unified DeepGRASP reconstruction model. An all-in-one DeepGrasp network can be trained once and then applied to perform GRASP reconstruction across different organs, spatial resolutions, and temporal frames, enabling a single model to handle a wide range of reconstruction tasks without task-specific retraining.

## Discussion and Future Directions

GRASP has demonstrated remarkable versatility and impact across a wide range of imaging applications. Its flexibility in data acquisition, robustness to motion, and compatibility with advanced reconstruction methods have made it a powerful framework in both research and clinical settings. This section briefly outlines the current limitations of the GRASP framework and highlights emerging directions being actively explored by the NYU team to further expand its utility.

### GRASP MRI with Different Sampling Trajectories

While radial sampling offers several key advantages, including intrinsic motion robustness, flexible data sorting, and the feasibility of self-navigation, it is not a one-size-fits-all trajectory for all clinical applications. For example, radial sampling is less suited than other trajectories for T2-weighted imaging and diffusion MRI. Moreover, with the renewed interest in low-field MRI, where improved B0 homogeneity mitigates limitations that typically hinder the broad use of non-Cartesian sampling with long readouts at high field (e.g., spiral sampling), it has become evident that radial trajectories may not always be the optimal choice at reduced field strengths (114).

To address these limitations, extending GRASP MRI beyond radial sampling is an important future direction. The NYU team has recently begun exploring spiral sampling, whose high SNR and efficient coverage may be advantageous for low-field applications, as well as PROPELLER sampling for T2-weighted and diffusion-weighted imaging. These multifaceted sampling schemes aim to retain the strengths of GRASP while improving image quality and efficiency in applications where radial sampling may fall short. Preliminary results in these directions have been encouraging as reported in several conference abstracts (115–119).

**Further Extension of Deep Learning-Based GRASP**

Building upon the initial development of DeepGrasp, future directions in the era of artificial intelligence (AI) are expected to focus increasingly on deep learning-based frameworks that move beyond pure reconstruction. One promising direction includes joint MRI reconstruction and denoising, in which neural networks simultaneously recover undersampled data and suppress noise. This could be particularly valuable for low-field MRI and applications requiring high spatial resolution. Another emerging direction is joint reconstruction and quantification with MP-GRASP, where MRI parameter estimation (e.g., T1, T2, perfusion and others) is directly integrated into the reconstruction process rather than performed as a separate step, as demonstrated in early studies (120–123). Such approaches hold the potential to further shorten scan times, reduce variability across patients and imaging sites, and accelerate the standardization of quantitative MRI for broader clinical implementation.

### Integration with MRI-Guided Therapy

Another exciting direction for GRASP MRI lies in the use for image-guided therapy, particularly MRI-guided radiation therapy. In fact, the XD-GRASP technique has already been well recognized in the radiation therapy field (124) and adopted in several abdominal cancer treatment protocols on MRI-Linac systems, where it provides a robust capability for respiratory-resolved 4D imaging to support treatment planning and motion management (125). More recently, novel extensions such as Live-View GRASP and similar techniques (110–112) have been proposed to overcome the latency bottlenecks of conventional 3D MRI by separating the imaging workflow into an off-line learning stage and a live-view stage. This enables near-instantaneous 3D image updates during treatment to facilitate high-fidelity motion tracking and adaptive response during radiation delivery.

These innovations expand the role of MRI beyond traditional diagnostics towards image-guided therapy. Future developments are expected to focus on deeper integration of GRASP MRI with treatment planning systems, real-time motion tracking, and adaptive dose delivery. However, successful clinical translation of these methods will depend on close collaboration between academic researchers and industry partners to ensure robust implementation and rigorous validation.

## Conclusion

This article summarized a decade of work from the NYU team in developing, translating, disseminating, and extending the GRASP MRI technique. It also described the historical context in which GRASP was conceived, building upon prior innovations in radial MRI, compressed sensing reconstruction, and dynamic imaging. From the initial implementation as a motion-robust technique for free-breathing DCE-MRI, GRASP has evolved into a versatile framework that supports a wide range of acquisition and reconstruction strategies, as summarized in this article. Looking ahead, the future of GRASP lies in further developing its capabilities across diverse sampling trajectories and field strengths, as well as in further integration with therapeutic applications, real-time imaging strategies, and intelligent reconstruction methods. With continued innovation and clinical translation, GRASP MRI may be expected to play an important role in further improving patient care in times to come.

## Financial Disclosure

The authors of this article have patents on the GRASP and XD-GRASP techniques.